\documentclass[conference]{IEEEtran}
\IEEEoverridecommandlockouts
\usepackage[noadjust]{cite}
\usepackage{amsmath,amssymb,amsfonts}
\usepackage{algorithm}
\usepackage{algorithmic}
\usepackage{amsthm}
\theoremstyle{definition}
\newtheorem{definition}{Definition}
\usepackage{graphicx}
\usepackage{textcomp}
\usepackage{xcolor}
\def\BibTeX{{\rm B\kern-.05em{\sc i\kern-.025em b}\kern-.08em
    T\kern-.1667em\lower.7ex\hbox{E}\kern-.125emX}}
\begin{document}

\title{Lifelong Learning for Minimizing Age of Information in Internet of Things Networks\\
\thanks{This work was supported by the National Natural Science Foundation of China (61971066) and the Beijing Natural Science Foundation (No. L182038), National Youth Top-notch Talent Support Program.}
}


\author{
	\IEEEauthorblockN{Zhenzhen Gong\IEEEauthorrefmark{2}, Qimei Cui\IEEEauthorrefmark{2}, Christina Chaccour\IEEEauthorrefmark{3}, Bo Zhou\IEEEauthorrefmark{3}, Mingzhe Chen\IEEEauthorrefmark{4}, and Walid Saad\IEEEauthorrefmark{3}}
	
	\IEEEauthorblockA{ \IEEEauthorrefmark{2} National Engineering Lab for Mobile Network Technologies, \\
	Beijing University of Posts and Telecommunications, Beijing, China. \\
		Emails: \{gongzhenzhen0822, cuiqimei\}@bupt.edu.cn.}
	\IEEEauthorblockA{\IEEEauthorrefmark{3} Wireless@VT, Bradley Department of Electrical and Computer Engineering, Virginia Tech, Blacksburg, VA, USA.  \\
		Emails: \{christinac, ecebo, walids\}@vt.edu.}
	\IEEEauthorblockA{ \IEEEauthorrefmark{4} Department of Electrical Engineering, Princeton University, Princeton, NJ, USA. Email: mingzhec@princeton.edu.}
}


\maketitle

\begin{abstract}
In this paper, a lifelong learning problem is studied for an Internet of Things (IoT) system.
In the considered model, each IoT device aims to balance its information freshness and energy consumption tradeoff by controlling its computational resource allocation at each time slot under dynamic environments.
An unmanned aerial vehicle (UAV) is deployed as a flying base station so as to enable the IoT devices to adapt to novel environments.
To this end, a new lifelong reinforcement learning algorithm, used by the UAV, is proposed in order to adapt the operation of the devices at each visit by the UAV.
By using the experience from previously visited devices and environments, the UAV can help devices adapt  faster to future states of their environment.
To do so, a knowledge base shared by all devices is maintained at the UAV. Simulation results show that the proposed algorithm can converge $25\%$ to $50\%$ faster than a policy gradient baseline algorithm that optimizes each device's decision making problem in isolation.

\end{abstract}


\section{Introduction}

The Internet of Things (IoT) \cite{zanella2014internet} is a promising technology that will enable a plethora of applications including smart metering systems, industrial control and autonomous robotics.
Normally, IoT devices are powered by batteries with limited capacity rather than a fixed power supply \cite{da2014internet}. Thus, the challenge of energy efficiency in IoT systems drew extensive attention recently.
Along with energy efficiency, in an IoT, it is also of fundamental importance to ensure the freshness of status information of the physical processes monitored by various IoT devices for accurate monitoring and control.
In order to quantify information freshness, the concept of \emph{age of information} (AoI) \cite{kaul2012real} has recently been proposed as a rigorous metric.

There have been extensive works that look at minimizing the AoI under various scenarios, including
IoT status update \cite{zhou2019joint},
energy harvesting systems \cite{wu2017optimal},
vehicular networks \cite{kaul2011minimizing},
cognitive networks \cite{wang2020minimizing}
and augmented reality services \cite{chaccour2020ruin}.
%
However, for some intelligent real-time IoT applications, the status freshness depends not only on the status update through the wireless channel, but also on the status data processing operations at the IoT devices.
For example, in smart video surveillance, the valuable information embedded in the images will not be available before some processing operations can be done \cite{li2020age}.
Different from energy-limited and small-sized IoT devices, unmanned aerial vehicles (UAVs) \cite{mozaffari2018beyond, chen2019artificial, chen2017caching}
have potentially more resources which can be used to complement the services of IoT networks,
particularly when deploying a ground base station can be too expensive or impossible such as in extreme environments, e.g., remote areas, primary forests, and congealed grounds.
Nevertheless, a UAV can easily fly to and communicate with IoT devices in a cost efficient way.
As such, in order to facilitate the IoT services while saving energy for IoT devices, one can exploit the role of UAV as a flying base station.

A number of existing studies have focused on key problems related to AoI optimization in UAV-based systems using machine learning methods.
In \cite{ferdowsi2020neural}, the authors study the UAV trajectory design problem by jointly considering the scheduling of status of update with the target of minimizing weighted sum-AoI.
The work in \cite{xu2020aoi} studies the dynamic status update strategy in caching enabled IoT networks with the reinforcement tool Sarsa.
In \cite{hatami2020age}, the age-aware status update for energy harvesting IoT systems is studied using classical Q-learning method.
Despite their promising results, these existing works do not consider the AoI optimization problem when the status update pattern varies over time. Normally, the physical surrounding environment is assumed to be stationary. However, the stationary assumption rarely holds in more realistic settings \cite{thrun1998lifelong}.
Under dynamically changing environments, it is important for IoT devices to adapt to new environments so as to optimize their AoI and energy consumption tradeoff.
Thus, there is a need for a novel framework that overcomes the aforementioned challenges of existing learning and AoI works.

The main contribution of this paper is, thus, a novel lifelong reinforcement learning approach for optimizing the AoI and energy consumption tradeoff of IoT devices while considering dynamic environments.
In the studied model, each IoT device collects and processes time-sensitive data packets from continuously changing environments.
By controlling the energy used at each time slot, each device must balance its AoI and energy tradeoff.
However, when the environment changes, the IoT devices must be able to adapt to the new environment while maintaining desirable AoI and energy levels.
To address this issue, a UAV is deployed as a flying base station to assist the devices
adapt to new environments quickly.
To do so, the UAV visits the ground devices and collects their interaction history, i.e., the information freshness and energy consumption dynamics generated at devices over time.
In order to enable the adaptation capability for the IoT devices under new environments,
we propose a novel lifelong reinforcement learning (LLRL) approach that allows the UAV to optimize each device's AoI-energy tradeoff by continuously accumulating past knowledge and using it in future learning processes.
Simulation results show that, after being trained, our approach can achieve lower AoI with less energy consumption
than a random initial policy when the environment begins to change.
Moreover, a faster convergence speed can be obtained compared to a policy gradient baseline algorithm.

\section{System Model And Problem Formulation} \label{SystemModel}
Consider an IoT network in which a set $\mathcal{N}$ of $N$ devices are randomly deployed to collect and process environment related data.
For example, consider an automated farm, in which IoT devices are installed to
detect environment events and take pictures to capture a given event.
After processing the captured images, the IoT devices can extract meaningful information such as the intrusion of an animal, the health status of plants, or any other useful event on the farm grounds.
In response to detected events, each IoT device must take an action (e.g., send an alarm or water the plants, etc.).
The frequency and processing requirements of such events can vary over time due to factors such as weather or human activity.
As can be seen above, sending an alarm or watering the plants can be time-sensitive. Over time, the data freshness will obviously decrease.
Though allocating more CPU resources can maintain the freshness of information, it comes at the expense of additional energy consumption.
%
Clearly, this process gives rise to a tradeoff between information freshness and energy consumption.

\subsection{System Model} \label{SingleDevice}
We consider a discrete-time system with slot $t=0,1,\ldots$. The environment-related data arrives to each device $i \in \mathcal{N}$  at the beginning of each time slot. The packets arrive at different devices are independent of each other and also independent and identically distributed (i.i.d.) over slots, following a Bernoulli distribution.
Let $p_{i,t} \in \{0,1\}$ indicate the arrival of a new data packet to device $i$ at the beginning of time slot $t$. As such $p_{i,t} = 1$ will be the arrival of a new data packet and $p_{i,t} = 0$ will indicate the absence of any new data arrival. The probability for Bernoulli distribution will be denoted as $\lambda_i$.
The CPU of each IoT device uses a first come first served (FCFS) scheduling  policy.
We define $u_{i,k}$ and $v_{i,k}$ as, respectively, the arrival and processing ending times for data packet $k$. $W_{i,k}$ and $S_{i,k}$ are the waiting time and the service time of packet $k$ at device $i$.


%

The energy consumption per CPU cycle is proportional to the square of CPU frequency \cite{CPU_energy}. Given the constant length of a time slot,
the energy consumption per time slot is equal to $\alpha_i \epsilon_{i,t}^3$, where $\epsilon_{i,t}$ is the number of CPU cycles at time slot $t$ and $\alpha_i$ is a constant related to device chip architecture.
We define $a_{i,t}$ as the size of a data packet that arrives at device $i$ at time slot $t$. For simplicity, it can be characterized by the number of CPU cycles required to process it.
Let $\bar{a}_i = \mathbb{E}[a_{i,t}]$ be the average packet size for device $i$.
At the beginning of each slot, the number of CPU cycles required to process the remaining data packets is denoted by $b_{i,t}$.
The queuing dynamics are given by: $b_{i,t+1} = \max \{b_{i,t} + p_{i,t}  a_{i,t} - \epsilon_{i,t}, 0 \}.$

\begin{figure}[htbp]
	\centering
	\includegraphics[width=3.5 in]{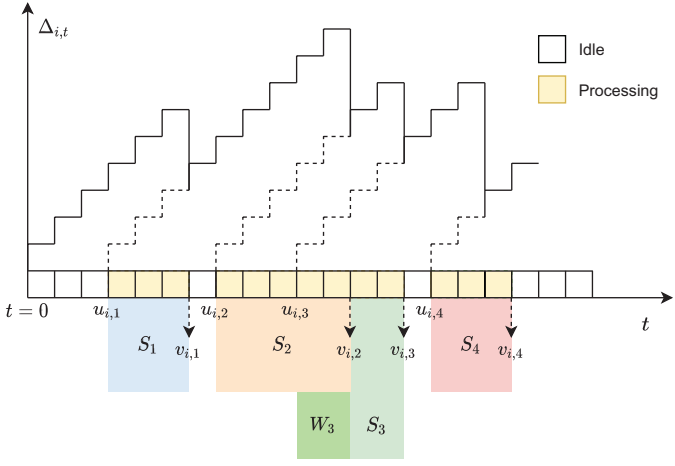}
	\caption{The dynamic evolution of AoI for device $i$. }
	\label{AoI_Sawtooth}
\end{figure}

We adopt the AoI as a performance metric to quantify the freshness of processed information \cite{li2020age}.
Formally, the AoI is defined as the time elapsed since the generation of the last received status update packet \cite{kaul2012real}.
Let $\Delta_{i,t}$ be the \emph{AoI} at device $i$ at the beginning of time slot $t$, such as $\Delta_{i,t} = t - u_{i,\xi_i}$,
where $\xi_i$ is the index of the latest processed data packet for device $i$ and $u_{i,k}$ represents the arrival time for data packet $k$. It is assumed that each IoT device collects each data packet without any additional delay.
The evolution of the AoI for a device $i$ is shown in Fig.~\ref{AoI_Sawtooth}. 
Accordingly, $\Delta_{i,t} \in \mathcal{N}$ can be updated as follows:
\begin{equation}
\setlength{\abovedisplayskip}{3pt}
\setlength{\belowdisplayskip}{4pt}
\Delta_{i,t+1} =
\begin{cases}
\Delta_{i,t} + 1 \, , & \textrm{if} \quad \omega_{i,t+1} = 0 \, , \\
(t+1)- u_{i,\xi_{i,t+1}} \, ,& \textrm{if} \quad \omega_{i,t+1} =1 \, ,\\
\end{cases}
\label{AoI update}
\end{equation}
where $\omega_{i,t+1} \in \{0,1\}$. $\omega_{i,t+1} =1$ implies the CPU cycles required for one data packet were all completed in the previous time slot, otherwise $\omega_{i,t+1} =0$. 

Given that the IoT devices need to operate in an energy-efficient manner, we need to characterize the \emph{freshness-energy} tradeoff during their interaction with their environment at each time slot. Hence, we define a cost function that captures this tradeoff, as follows:
\begin{equation}
c_i(t)  =  \beta \Delta_{i,t} + (1-\beta) \alpha_i \epsilon_{i,t}^3 \, ,
\label{cost_function}
\end{equation}
where $\beta \in [0,1]$ is a weighting parameter to balance AoI and energy efficiency.
For all the devices in the IoT system, the overall AoI minimization problem can be characterized as:
\begin{alignat}{2}
\min_{\Pi} & \quad  \lim_{T\to\infty} \dfrac{1}{T}  \sum_{t=0}^{T}  \sum_{i=1}^{N}  \beta \Delta_{i,t} + (1-\beta) \alpha_i \epsilon_{i,t}^3 \label{Problem_1} \\
\mathrm{s.t.} \quad
& \Delta_{i,t} \in \mathbb{N} \quad  \forall i= 1, 2, \ldots, N, \forall t \in \mathbb{N}  \, , \nonumber \\
&0 \le \epsilon_{i,t} \le \epsilon_{i,\textrm{max}},  \quad \forall i= 1, 2, \ldots, N, \forall t \in \mathbb{N}  \, ,            \nonumber \\
d_{i,t}  \in & \{0,1\},  \omega_{i,t} \in \{0,1\} \quad \forall i= 1, 2, \ldots, N, \forall t \in \mathbb{N}  \, , \nonumber
\end{alignat}
where $\Pi = \{ \Pi_1, \ldots, \Pi_N\}$ is the set of interaction policies for all devices. 
$\Pi_i$ is the interaction policy for device $i$.


The \emph{interaction policy} is thus defined as the policy using which, each device decides on how to use its CPU cycles at each slot in order to balance its AoI and energy tradeoff.
However, in practice, the environment of the IoT devices is often non-stationary.
For instance, in the previously discussed example of an automated managed farm, the environment changes can happen either naturally or artificially.
As a result, a policy that is derived for a given environment state may not apply to another environment state.
This requires each device to continuously adapt its interaction policy.

\vspace{0.5em}

To assist all the IoT devices with their interaction policy optimization, a central agent could be needed.
However, for battery limited and small-sized IoT devices, it is energy consuming to communicate with a remote base station when compared to a UAV that can fly closer to devices.
To do so, the UAV must visit the ground IoT devices one after another repeatedly.  
%
%

\vspace{0.5em}

We now observe that solving problem \eqref{Problem_1} using conventional optimization methods can be very complex.
On the one hand, awareness of future environmental changes is not possible to obtain.
On the other hand, the network will not have the distribution of the AoI, nor of the CPU cycle dynamics over time.
As such, the problem cannot be solved using standard tools such as convex optimization.
Next, we will use the parameterized interaction policy to enable the policy based optimization, which has no requirements on AoI or CPU cycles dynamics distribution.
After that, problem \eqref{Problem_1} will be reformulated and optimized over a parameterized interaction policy.

\subsection{Problem Formulation} \label{AlgoPGELLA}
In our model, the non-stationary environments experienced by the IoT devices will be modeled as a series of independent tasks. Each task is essentially a Markov decision process (MDP).
\begin{definition}
	For a single device, we define a task $\emph{Z}_{j} = ( \lambda_{j},  \bar{a}_{j}, \alpha_{j}, \epsilon_{j,\textrm{max}})$, where $j \in \mathbb{N}$, as a four-dimensional tuple that captures environment state $j$ of the device.
	The parameters in $\emph{Z}_{j}$ represent the  arrival rate, averaged data packet size, chip type of device, and maximum computing capability of the device.
%
	%
	\label{Definition_Task_Tuple}
\end{definition}


Given that all the IoT devices in the system experience i.i.d. environment changes, the set of tasks on one device is identical to all other devices. As such, without loss of generality, we let $j$ be the index of all the tasks and let each task start with $t=0$.
An interaction history $\boldsymbol{\tau}$, also called dynamic status updates, will be generated at each IoT during its interaction with environment.
In particular, a \emph{task-specific trajectory} $\boldsymbol{\tau}$ can be defined as a set of dynamic status updates produced under a task-specific environment state.
The dynamic space of a trajectory can be given by $\{\mathcal{X}, \mathcal{Y}, \mathcal{R}\}$ where $\mathcal{X}$ is the state space, $\mathcal{Y}$ is the action space, and $\mathcal{R}$ is the reward function. The state space $\mathcal{X}  \subset \mathbb{R}^d$ is the set of AoI values and number of pending data cycles at the beginning of each time slot, i.e., $\mathcal{X} = \{\boldsymbol{x}_{j,t}\}=  \{  \Delta_{j,t}, b_{j,t}  | \Delta_{j,t} \in \mathbb{N}, b_{j,t} \in  \mathbb{N}\}$. Here, $d$ equals to the number of variables in state space. The action space is the set of all possible CPU cycles,  $\mathcal{Y} = \{\boldsymbol{y}_{j,t}\} = \{\epsilon_{j,t}| 0 \le \epsilon_{j,t} \le \epsilon_{j,\textrm{max}} ,  \epsilon_{j,t}  \in \mathbb{R}\}$.
The reward function can be defined as $R(\boldsymbol{x}_{j,t}, \boldsymbol{y}_{j,t}) = - (\beta \Delta_{j,t} + (1-\beta) \alpha_j \epsilon_{j,t}^3  )$, which is the negative of the objective function \eqref{Problem_1}.
The parameterized policies can be defined as $\Pi'_j = \{ \pi_{{\boldsymbol{\theta}}_j} | \boldsymbol{\theta}_j \in \mathbb{R}^d \}$, where $\pi_{{\boldsymbol{\theta}}_j}(\boldsymbol{y}_{j,t} | \boldsymbol{x}_{j,t}) = \text{Pr}\{ \boldsymbol{y}_{j,t} | \boldsymbol{x}_{j,t}, \boldsymbol{\theta}_j  \} $.


The exact task tuple for a given environment is not known to neither the UAV nor the devices.
Thus, it is necessary for the UAV to infer a task using the collected trajectory.
Given a trajectory $\{\boldsymbol{x}_{j,t}, \boldsymbol{y}_{j,t}, R(\boldsymbol{x}_{j,t}, \boldsymbol{y}_{j,t})\}$, where
$0 \le t \le T_j$ and $T_j$ is the length of the trajectory, let $\mathcal{Q}_j = \{t \in \mathbb{N}| b_{j,t+1} +\epsilon_{j,t} >  b_{j,t}\}$ be the set of time slots at which there was a new packet arrival.
Hereinafter, the task-specific tuple can be identified as: $\lambda_j \approx T_j/Q_j$ and $\bar{a}_j \approx 1/Q_j \sum_t (b_{j,t+1} - b_{j,t})$, where $t \in \mathcal{Q}_j$.
Since $\alpha_j$ and $\epsilon_{j,\textrm{max}}$ are related to the IoT device type, they can be directly obtained from the device by the UAV.
As such, the task tuple $( \lambda_j,  \bar{a}_j, \alpha_j, \epsilon_{j,\textrm{max}})$ can be obtained from the trajectory which is generated under the task specific environment.
Hereinafter, this process will be called a \emph{task discovery process}.

Our goal is to achieve a balance between AoI and energy consumption for all the devices.
Given that all the IoT devices are experiencing i.i.d. environment changes, problem \eqref{Problem_1} can be transformed into an optimization problem whose objective is to find the policy that maximizes the expected average return for all tasks:
%
\begin{equation}
\setlength{\abovedisplayskip}{2pt}
\setlength{\belowdisplayskip}{3pt}
\max_{\Pi'}  \quad  \lim_{M\to\infty} \dfrac{1}{M}  \sum_{j=1}^{M} \mathcal{J}(\boldsymbol{\theta}_j)
\quad  \mathrm{s.t.} \,
\mathcal{X} \subset \mathbb{R}^d, \mathcal{Y},\mathcal{R} \subset \mathbb{R} \, ,
\label{goal_final}
\end{equation}
where $\mathcal{J}(\boldsymbol{\theta}_j) = \int p_{\boldsymbol{\theta}_j}(\boldsymbol{\tau}) \mathfrak{R_j}(\boldsymbol{\tau}) \textrm{d} \boldsymbol{\tau}$ and
$M$ is the number of tasks observed by the UAV.
$\Pi'$ is the set of policies for all tasks.
Instead of optimizing over all potential tasks, the number of tasks observed so far will be used.
$p_{\boldsymbol{\theta}_j}$ represents the probability distribution for trajectory $\boldsymbol{\tau}$ and $\mathfrak{R_j}(\boldsymbol{\tau})$ is the gain for a given trajectory. In other words, we have:
\begin{equation}
\setlength{\abovedisplayskip}{3pt}
\setlength{\belowdisplayskip}{3pt}
\setlength{\abovedisplayskip}{3pt}
\setlength{\belowdisplayskip}{3pt}
p_{\boldsymbol{\theta}_j}(\boldsymbol{\tau}) = P_0(\boldsymbol{x_0}) \prod_{t = 0}^{T} p(\boldsymbol{x}_{j,t+1}| \boldsymbol{x}_{j,t}, \boldsymbol{y}_{j,t}  )\, \pi_{{\boldsymbol{\theta}}_j}(\boldsymbol{y}_{j,t} | \boldsymbol{x}_{j,t}) \, ,
\end{equation}
\begin{equation}
\setlength{\abovedisplayskip}{3pt}
\setlength{\belowdisplayskip}{3pt}
\mathfrak{R_j}(\boldsymbol{\tau}) = \dfrac{1}{T} \sum_{t = 0}^{T} \gamma_j^{t-1}  \, R(\boldsymbol{x}_{j,t}, \boldsymbol{y}_{j,t})  \, ,
\end{equation}
where
$p(\boldsymbol{x}_{j,t+1}| \boldsymbol{x}_{j,t}, \boldsymbol{y}_{j,t}  )$ is the unknown state transition probability that maps a state-action pair at time slot $t$ onto a distribution of states at time slot $t+1$.
%

%



Traditional reinforcement learning (RL) algorithms such as Q-learning and its variants cannot be used to solve problem \eqref{goal_final} because their ability to learn is limited to a specific stationary environment and their performance will be reduced when it comes to the environment changing situation.
As such, next, to solve \eqref{goal_final}, a lifelong reinforcement learning algorithm \cite{chen2018lifelong} for our IoT system is proposed so as to enable long-term adaptation to dynamic environments.
%

\section{Lifelong Reinforcement Learning} \label{LLRLChapter}
Now, we introduce our lifelong reinforcement learning algorithm that merges the concept of knowledge transfer with the policy gradient (PG) framework.
As a classical RL algorithm, PG can find the optimal interaction policy for a single task by running gradient descent over a policy space towards the maximal average reward without any requirements on the state transition probability.
However, this process is fitted only to certain environment states, which can lead to poor performance in non-stationary environments.
To address these challenges, we propose a lifelong learning approach that can enable knowledge transfer between past and future tasks thus enabling the devices adapt to the dynamic and unknown environments.

To enable knowledge transfer between tasks, we assume that the policy parameters for each task can be a linear combination of these $h$ latent components \cite{kumar2012learning},
such that $\boldsymbol{\theta}_j = \boldsymbol{L}\boldsymbol{s}_j$, where $\boldsymbol{s}_j \in \mathbb{R}^{h}$ is a vector of linear parameters and $\boldsymbol{L}$ is a knowledge base with a library of $h$ latent components. The dimension of $h$ is chosen independently via cross-validation.
In order to maximize the knowledge captured by the latent components, the mapping function should be sparse. As such, each observed task can be a linear combination of only a few latent components in $\boldsymbol{L}$. With knowledge transfer considered, a lifelong learning loss function can be defined as :
\begin{equation}
\setlength{\abovedisplayskip}{3pt}
\setlength{\belowdisplayskip}{3pt}
e_\textrm{T} (\boldsymbol{L})  = \dfrac{1}{M} \sum_{j = 1}^{M} \min_{\boldsymbol{s_j}} \Big[ -\mathcal{J}(\boldsymbol{\theta}_j) + \eta_1  \lVert \boldsymbol{s_j}  \lVert_{1} \Big]  + \eta_2 \lVert \boldsymbol{L}  \lVert_{\textrm{F}}^2 \, ,
\label{Target_1}
\end{equation}
where $L_1$ norm approximates the true vector sparsity and $\lVert \boldsymbol{L} \lVert_{\textrm{F}} = (tr(\boldsymbol{L}\boldsymbol{L}'))^{1/2}$ is the Frobenius norm of matrix $\boldsymbol{L}$. Parameter $\eta_1$ controls the sparsity of $\boldsymbol{s}_j$. The penalty on the Frobenius norm of $\boldsymbol{L}$ regularizes the predictor weights to have low $\ell_2$ norm and avoids overfitting.



\begin{algorithm}
	\caption{Lifelong Reinforcement Learning (LLRL)}
	\label{PG-ELLAAlgo}
	\begin{algorithmic}
		\REQUIRE $T \leftarrow 0$, $\boldsymbol{A} \leftarrow \mathrm{zeros}_{d \times h, d \times h}$,
		\REQUIRE $\boldsymbol{b} \leftarrow \mathrm{zeros}_{d \times h, 1}$, $\boldsymbol{L} \leftarrow \mathrm{zeros}_{d, h}$
		\WHILE{some device $i$ is available}
		\STATE $\boldsymbol{\tau} \leftarrow \mathrm{CollectTrajectorySinceLastVisit()}$
		\STATE Identify task-specific tuple $\{\lambda,  \bar{a}, \alpha, \epsilon_{\textrm{max}}\}$ for device $i$
		\IF{The same as the previous task}
		\STATE $\boldsymbol{A} \leftarrow \boldsymbol{A} - \big( \boldsymbol{s} \boldsymbol{s}^\mathrm{T}  \big) \otimes \boldsymbol{\Gamma} $
		\STATE $\boldsymbol{b} \leftarrow \boldsymbol{b} - \mathrm{vec} \big( \boldsymbol{s}^\mathrm{T}  \otimes \big(   \boldsymbol{\alpha}^\mathrm{T} \boldsymbol{\Gamma}  \big) \big)$
		\ELSE
		\STATE Identify a new task $j$ for device $i$
		\STATE $M \leftarrow M+1$
		\ENDIF
		\STATE Compute $\boldsymbol{\alpha}_j$  and $\boldsymbol{\Gamma}_i$ from $\boldsymbol{\tau} $
		\STATE $\boldsymbol{L} \leftarrow \mathrm{reinitializeAllZeroColumns}(\boldsymbol{L})$
		\STATE $\boldsymbol{s} \leftarrow \arg \min_{\boldsymbol{s}} \ell \big(  \boldsymbol{L},  \boldsymbol{s}, \boldsymbol{\alpha}, \boldsymbol{\Gamma} \big)$
		\STATE $\boldsymbol{A} \leftarrow \boldsymbol{A} + \big( \boldsymbol{s} \boldsymbol{s}^\mathrm{T}  \big) \otimes \boldsymbol{\Gamma} $
		\STATE $\boldsymbol{b} \leftarrow \boldsymbol{b} + \mathrm{vec} \big( \boldsymbol{s}^\mathrm{T}  \otimes \big(   \boldsymbol{\alpha}^\mathrm{T} \Gamma  \big) \big)$
		\STATE $\boldsymbol{L} \leftarrow \mathrm{mat} \Big(  \big( \frac{1}{M} \boldsymbol{A} + \eta_2 \boldsymbol{I}_{d \times h, d \times h} \big) ^{-1} \frac{1}{M} \boldsymbol{b} \Big) $
		\ENDWHILE
	\end{algorithmic}
\end{algorithm}

\subsection{Eliminating Dependencies}
To solve \eqref{Target_1}, an exhaustive evaluation of all the tasks is required.
However, the UAV can only fly to one device at its visit to the device, without knowing what other tasks are.
To eliminate the dependency on other tasks,
%
%
%
%
%
%
%
%
%
%
a second-order Taylor expansion can be done to approximate $-\mathcal{J}(\boldsymbol{\theta}_j)$. Thus, the approximation of $e_\textrm{T} (\boldsymbol{L})$ can be given as:
\begin{equation}
\hat{e}_\textrm{T} (\boldsymbol{L})  =
\dfrac{1}{M} \sum_{j = 1}^{M} \min_{\boldsymbol{s}_j}
\Bigg[
\displaystyle{ \lVert  \boldsymbol{\alpha}_j - L \boldsymbol{s}_j  \lVert^2_{\boldsymbol{\Gamma}_j} }
+ \eta_1  \lVert \boldsymbol{s}_j  \lVert_{1}
\Bigg]
+ \eta_2 \lVert \boldsymbol{L}  \lVert_{\textrm{F}}^2 \, ,
\label{Target_final}
\end{equation}
where $\lVert \boldsymbol{v} \lVert^2_{\boldsymbol{A}} = \boldsymbol{v}^\mathrm{T}\boldsymbol{A}\boldsymbol{v}$,
$\boldsymbol{\alpha}_j$ is the optimal policy parameter that can be obtained by any PG algorithm and
$\boldsymbol{\Gamma}_j$ is the Hessian matrix \cite{ammar2014online}. 

From \eqref{Target_final}, we can see that each evaluation of  the latent basis $\boldsymbol{L}$ will require a minimization over all mapping functions $\boldsymbol{s}_j$ of all tasks, which can complicate the learning process when the number of tasks grows. To address this issue, we fix the mapping functions of other tasks, when the mapping function of a current task is optimized.
This is a reasonable approach because any change to the policy can be obtained from the update of the latent basis $\boldsymbol{L}$. We next explain this method in more detail.

\begin{figure}[htbp]
	\centerline
	{\includegraphics[width=3.5 in]{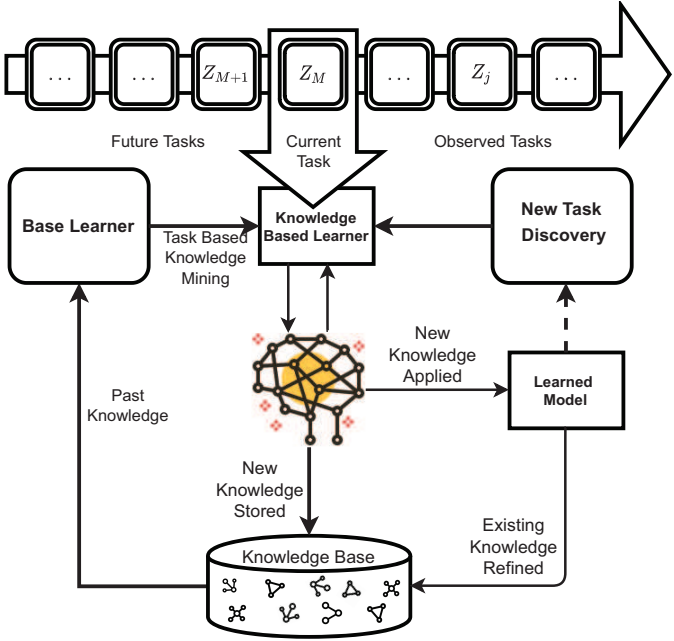}}
	\caption{Diagram of the proposed lifelong reinforcement Learning algorithm.}
	\label{Flow}
\end{figure}

With the above mentioned simplification, $\boldsymbol{s}_j$ and $\boldsymbol{L}$ in equation \eqref{Target_final} can be updated recursively, as follows:
\begin{equation}
\setlength{\abovedisplayskip}{3.8pt}
\setlength{\belowdisplayskip}{3pt}
\boldsymbol{s}_j \leftarrow \arg \min_{\boldsymbol{s}_j} \ell \big(  \boldsymbol{L},  \boldsymbol{s}_j, \boldsymbol{\alpha_j}, \Gamma_j \big) \, ,
\label{s}
\end{equation}
\begin{equation}
\setlength{\abovedisplayskip}{3pt}
\setlength{\belowdisplayskip}{2.5pt}
\boldsymbol{L} = \arg \textrm{min}_{\boldsymbol{L}} \dfrac{1}{T} \sum_{t=0}^{T} \ell \big(  \boldsymbol{\boldsymbol{L}},  \boldsymbol{s}_j, \boldsymbol{\alpha}_j, \boldsymbol{\Gamma}_j \big)
+
\eta_2 \lVert \boldsymbol{\boldsymbol{L}}  \lVert_{\textrm{F}}^2 \, ,
\label{L_update}
\end{equation}
where the loss function has the following form:
\begin{equation}
\ell \big(  \boldsymbol{L},  \boldsymbol{s}_j, \boldsymbol{\alpha}_j, \boldsymbol{\Gamma}_j \big) =
 \lVert  \boldsymbol{\alpha}_j - \boldsymbol{L} \boldsymbol{s}_i  \lVert^2_{\boldsymbol{\Gamma}_j}
+ \eta_1  \lVert \boldsymbol{s}_j  \lVert_{1}  \, ,
\end{equation}
hereinafter, \eqref{s} is a $L_1$-regularized regression problem which can be solved as an instance of Lasso.
The update of $\boldsymbol{L}$ can be obtained by $\boldsymbol{A}^{-1}\boldsymbol{b}$: 
\begin{equation}
\setlength{\abovedisplayskip}{3pt}
\setlength{\belowdisplayskip}{3pt}
\boldsymbol{A} = \eta_2 \boldsymbol{I}_{d \times h, d \times h} + \dfrac{1}{T} \sum_{t=0}^{T}
\big( \boldsymbol{s}_j \boldsymbol{s}_j^\mathrm{T}  \big)
   \otimes \boldsymbol{\Gamma}_j   \, ,
\label{A}
\end{equation}
\begin{equation}
\setlength{\abovedisplayskip}{3pt}
\setlength{\belowdisplayskip}{3pt}
\boldsymbol{b} = \dfrac{1}{T} \sum_{t=0}^{T}  \mathrm{vec}
\big( \boldsymbol{s}_j^\mathrm{T}  \otimes \big(   \boldsymbol{\alpha}_j^\mathrm{T} \boldsymbol{\Gamma}_j  \big) \big) \, .
\label{b}
\end{equation}
Thus, by computing $\boldsymbol{A}$, $\boldsymbol{b}$, and $\boldsymbol{s}_j$ incrementally, the dependency on all trajectories can be eliminated. As such, the update of $\boldsymbol{L}$ can be computed efficiently. The complete flow of the approach is presented in Algorithm 1. 

\subsection{Lifelong Learning Procedure}

The lifelong learning algorithm diagram for our UAV-IoT system from the perspective of knowledge learning is presented in Fig.~\ref{Flow}.
Detailed procedures are as below:
\begin{enumerate}
	\item \emph{Initialize policy at devices}: Random initialization policies are used at the very start.
	\item \emph{Collect trajectory}: The UAV will randomly choose one device as its destination and flies to it. Upon arrival, it collects the trajectory from the device.
	\item \emph{New task discovery}: The UAV uses the collected device trajectory to identify if it was experiencing a new environment since its last visit using the task discovery method described in Section \ref{AlgoPGELLA}.
	\item \emph{Knowledge mining for current task}: The task-specific knowledge contained in the Hessian matrix will be computed. Instead of looping until the policy parameters converge, only one run of the policy gradient update will be performed.
	The PG algorithm used to compute policy gradient will be called base learner.
	\item \emph{Refine knowledge base}: If the identified task is new, the knowledge base will be refined according to \eqref{A} and \eqref{b}. Otherwise,
	the outdated knowledge will be deducted before knowledge base refinement.
	\item \emph{Update model}: The parameterized policy can be updated as in \eqref{s} and \eqref{L_update}.
	\begin{figure}[htbp]
		\centerline{\includegraphics[width=3.7 in]{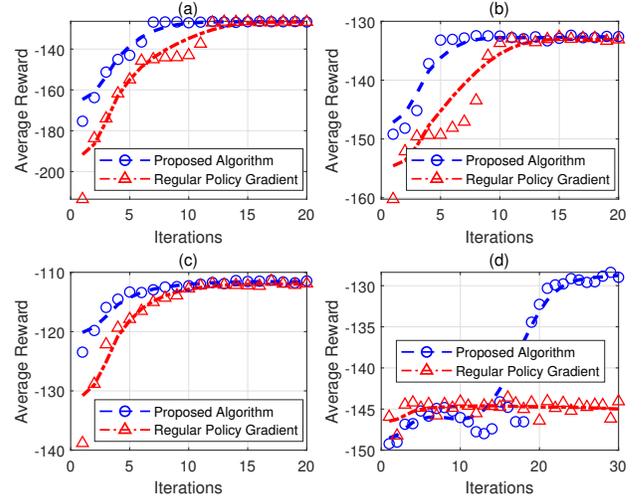}}
		\caption{The proposed algorithm provides better starting point and converges faster than regular policy gradient method.
		}
		\label{Test_Single}
	\end{figure}
	\item \emph{Transmit updated policy back to the device}: The updated policy for the current environment will be transmitted back to the device. The device will use it as the initial policy from now on before a new policy received from the UAV.
\end{enumerate}

As such, the UAV can provide adaptive capabilities to the devices based on the accumulated knowledge. In the training phase, the knowledge base will be trained using 10 tasks distributed over 3 devices.
During training, the same device will be visited by the UAV more than once.
A new task will appear to a device only when the policy parameters of the previous task have converged.
The training procedure ends when the policy parameters of all training tasks converge.

Finally, we show that our LLRL can support a variety of base learners.
In theory, any policy based reinforcement learning algorithm that can provide an estimate of the Hessian matrix can be incorporated.
Take REINFORCE \cite{williams1992simple} as an example.
With Gaussian policy $\pi_{{\boldsymbol{\theta}}_j}(\boldsymbol{y}_{j,t} | \boldsymbol{x}_{j,t}) \sim \mathcal{N}(\boldsymbol{\theta}_j^{\mathrm{T}} \boldsymbol{x}_j , \boldsymbol{\sigma}_j^2)$ considered, the first-order derivative has the following form:
%
$
-
\mathbb{E}
[
\mathfrak{R_j}(\boldsymbol{\tau})
(
\sum_{t=0}^{T}
\boldsymbol{\sigma}_j^{-2}\big( \boldsymbol{y}_{j,t}  - \boldsymbol{\theta}_j^{\mathrm{T}} \boldsymbol{x}_{j,t} \big) \boldsymbol{x}_{j,t}^{\mathrm{T}}
)
]$.
Thus, the Hessian matrix can be easily obtained as: $\boldsymbol{\Gamma}_j = \mathbb{E} \Big[ \sum_{t=0}^{T} \boldsymbol{\sigma}_j^{-2} \boldsymbol{x}_{j,t} \boldsymbol{x}_{j,t}^\mathrm{T}  \Big]
\label{Gamma}$.

\section{Simulation Results and Analysis} \label{Simulation}

\begin{figure}[htbp]
	\centerline{\includegraphics[width=3.7  in]{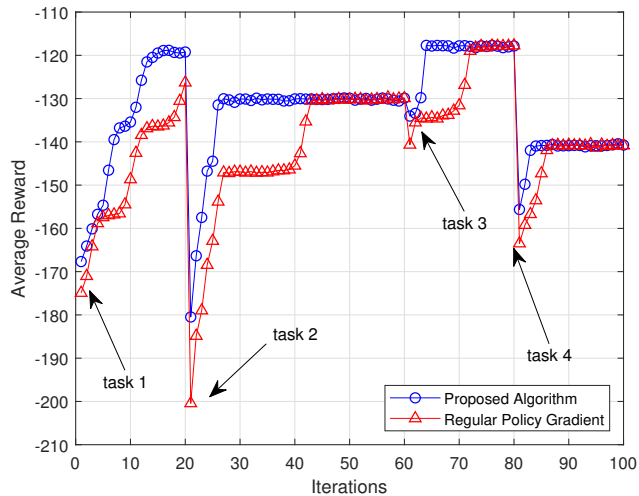}}
	\caption{A sequential learning process.}
	\label{Test_Long}
\end{figure}

For our simulations, we consider an actual Mica2 Platform \cite{hill2002mica}.
We adopt models with a CPU frequency between 315 MHz to 916 MHz. Each bit requires $10^5$ CPU cycles. The length of a time slot is $1$ second.
According to the processor parameters of the selected Mica2 models, $\alpha = 10^{-21}$ $\rm J/cycles^3$. We also define the following parameters: $\mathbb{E}[\bar{a}] \in [1 \times 10^7, 5 \times 10^7 ]$ cycles, $\textrm{Var}[\bar{a}] = 5 \times 10^6$ cycles, $\epsilon_{\textrm{max}} \in [3 \times 10^6,8 \times 10^6]$ cycles/slot. We set the weighted parameter $\beta = 0.5$ and arrival rates for different tasks $\lambda \in [1,5]$.
Each trajectory has 50 dynamic status updates and the reward is averaged over 100 trajectories.
According to cross validation, the value of $h$ is selected to be the same as the dimension of our task space.

After training, the UAV that acts as a learning agent is going to be tested with a set of independent new tasks.
Each test task is generated randomly as in training phase. In Fig.~\ref{Test_Single}, we analyze the learning performance of the LLRL approach and regular policy gradient algorithm over the same new task.
Clearly, our LLRL approach can provide a better starting policy for new tasks due to the knowledge accumulation.
Our LLRL approach yields a 10\% improvement on average reward at the beginning of a task compared to random initial policy.
Our method converges faster than the regular policy gradient method 50\% faster in the best case and 25\% faster in the worst case.
Four examples are presented as in Fig.~\ref{Test_Single}.
It is worth noting that in case (d), our method can achieve global optimum by leveraging the accumulated knowledge, while the baseline algorithm can only achieve the local optimum.

In Fig.~\ref{Test_Long}, we have inspected the lifelong learning capability using a series of new tasks.
The mean average reward over all iterations of our algorithm is 8\% higher than the mean average reward for regular policy gradient algorithm.
The average reward improvement can be achieved even when new tasks are emerging consecutively.
This implies the sustainability of our lifelong reinforcement learning algorithm.

\section{Conclusion} \label{Conclusion}
In this paper, we have studied the optimization of the AoI-energy tradeoff for IoT devices that are experiencing dynamic environments.
We have particularly considered a model in which a UAV acts as a central agent.
To solve this problem, we have developed a novel lifelong reinforcement learning algorithm that enables IoT devices to adapt to changing environments.
Simulation results have shown that the proposed LLRL algorithm can yield up to $50\%$ improvement in the convergence speed, compared to policy gradient baseline algorithm.
Future work can consider the trajectory design of the UAV.



\bibliographystyle{IEEEtran}
\bibliography{conference_101719}

\end{document}